\documentstyle[prl,aps,epsfig,eqsecnum,amssymb,amsmath,amsthm]{revtex}
\tighten
\newcounter{theo}
\newcounter{defi}
\newcounter{lemm}
\begin{document}
\clearpage
\preprint{}
\draft
\title{Some bounds for quantum copying with multiple copies}

\author{ A. E. Rastegin }
\address{Irkutsk State University, Irkutsk,
         664003, Russia \\
{\rm rast@api.isu.runnet.ru}}

\setcounter{page}{1}
\setcounter{theo}{0}
\setcounter{defi}{0}
\setcounter{lemm}{0}
\maketitle
\begin{abstract}
We study the relative error of the state-dependent $N\to L$
cloning. A copying transformation and dimension of state space are
not specified. Only the unitarity of quantum mechanical
transformations is used. The proposed approach is based on the notion
of the angle between two states. Firstly, the notion of the angle
between two states is discussed. The lower bound on the relative
error of copying with multiple copies is examined. In addition, the
lower bound on the absolute error is then studied. We compare
the obtained bounds with the case of maximizing the global fidelity.
\end{abstract}

%\date{\today}
\pacs{2001 PACS numbers: 03.65.Ta, 03.67.-a}
%\widetext

\pagenumbering{arabic}
\setcounter{page}{1}

\section{Introduction}

A copying of the quantum information has some severe constraints.
First of all, an arbitrary quantum state cannot be perfectly copied
\cite{wootters,barnum}. For copying of set of two non-orthogonal
pure states some bounds already appear. The two-state problem was
first considered by Hillery and Bu\v{z}ek \cite{hillery}.
They examined approximate cloning machines destined
for copying of prescribed two non-orthogonal states. In paper
\cite{brass} such devices were called 'state-dependent cloners'.
Writers of \cite{brass} introduced the notion of 'global fidelity'
and constructed the optimal symmetric state-dependent cloner which
optimizes the global fidelity. The problem is stated in the following
way. Let us assume that our our auxiliary device CM (the copying
machine) must produce two copies (one actual copy plus the original)
of particle secretly prepared in some state from a set
${\mathfrak{A}}=\{|\phi\rangle,|\psi\rangle\}$.  How well CM can do?
If states $|\phi\rangle$ and $|\psi\rangle$ are not orthogonal, then
errors will be inevitably introduced. The general lower bound on the
absolute error of copying of two-state set was obtained in paper
\cite{hillery}. Hillery and Bu\v{z}ek also considered the lower bound
on the absolute error in the case, when CM produces $n+1$ copies
(i.e. $n$ actual copies plus the original) of single input particle.
These results were extended in papers \cite{chefles,macchi} those
examined $N\to L$ state-dependent cloning. (Note that pointed works
also concern another questions.) The universal copying machines,
producing multiple copies, were considered in papers
\cite{werner,zanardi}.

Of course, the evaluation of copying quality is dependent on the
used measure of 'closeness' to ideality. Measure used by Hillery and
Bu\v{z}ek can be named 'absolute error' of copying of two-state set.
In work \cite{brass,chefles,macchi} the global fidelity was maximized
and the optimal symmetric state-dependent cloner was constructed.

Thus, the state-dependent cloning was mostly examined from the
'global fidelity' viewpoint. However, the state-dependent cloning is
a complex subject with many facets. Important as the notion of the
global fidelity is, it does not cover the problem on the whole. In
paper \cite{rastegin} author proposed and motived the notion of the
relative error for $1\to2$ state-dependent cloning. The lower bound
on the relative error was deduced. Our approach based on the notion
of the angle between two states. In present paper the developed us
approach is applied to the copying machine, giving a fixed number $L$
of copies from a fixed number $N$ of identically prepared particles
(it is clear, $L>N$). In submitted analysis a copying transformation
and dimension of state space are not specified. Only the unitarity of
quantum mechanical transformation is used. The lower bound on the
relative error is derived. The lower bound on the relative error
allows to elucidate a tradeoff between a quality of copies and
the part of actual copies. We also improve results obtained by
Hillery and Bu\v{z}ek \cite{hillery}. We describe the optimal
asymmetric state-dependent cloner which minimizes both the relative
and absolute errors. In our  examination all the state vectors are
normalized to unity. In calculations non-unit vectors will sometimes
occur, and these cases will be expressly stated. The norm of the
vector $|\Phi\rangle$ is defined as
$\,\|\:\!|\Phi\rangle\|=[\langle\Phi|\Phi\rangle]^{1/2}\,$.

\protect\section{Preliminary}

We shall now discuss the notion of angle between two states.
Angle $\:\delta(\Phi,\Psi)\in[0;\pi/2]\:$ between two
states $|\Phi\rangle$ and $|\Psi\rangle$ is defined by
\begin{equation}
\delta(\Phi,\Psi)
\ {\mathop=^{{\rm def}}} \
\arccos\left(
\bigl| \langle\Phi | \Psi\rangle \bigr| \right) \!\ .
\end{equation}
For brevity we shall also often write $\delta_{\Phi\Psi}\,$.
In paper \cite{rastegin} the following useful inequality was proven:
\begin{equation}
\bigl|
\langle\Phi| \,\Pi\, |\Phi\rangle -
\langle\Psi| \,\Pi\, |\Psi\rangle
\bigr|\leq
\sqrt{1-\bigl|\langle\Phi|\Psi\rangle\bigr|^2}
=\sin\delta_{\Phi\Psi}
\!\ ,
\label{ineq1}
\end{equation}
where $\Pi$ is any projector. So that if angle $\delta_{\Phi\Psi}$ is
small, then the probability distributions generated by states
$|\Phi\rangle$ and $|\Psi\rangle$ for an arbitrary measurement are
close to each other:
\begin{equation}
\bigl| P(R\,|\,\Phi) -
P(R\,|\,\Psi) \bigr| \leq \sin\delta_{\Phi\Psi} \!\ .
\label{ineqprob}
\end{equation}
It must be stressed that for mixed states there is a similar
relation. Suppose that $\chi$ and $\omega$ are density
operators describing states of a quantum system I. We can imagine
that these mixed states arise by a partial trace operation from pure
states of an extended system "I+I{\thinspace}I". That is, there are
states $|u\rangle$ and $|v\rangle$, for which
\begin{equation*}
\chi={\rm Tr}_{\scriptscriptstyle{\rm II}}
[\, |u\rangle\langle u| \,]
\ \  {\rm and} \ \
\omega={\rm Tr}_{\scriptscriptstyle{\rm II}}
[\, |v\rangle\langle v| \,]
\!\ .
\end{equation*}
These pure states $|u\rangle$ and $|v\rangle$ are called
"purifications" of $\chi$ and $\omega$ respectively. Fidelity
\begin{equation}
F(\chi,\omega)
\ {\mathop=^{{\rm def}}} \
\sup \bigl\{ \
\bigl|\langle u|v \rangle\bigr|^2
 \ \big| \
|u\rangle \ and \ |v\rangle \ are \ purifications \
of \ \chi \ and \ \omega \ \bigr\}
\label{fiddef}
\end{equation}
was introduced by Jozsa \cite{jozsa}. Fidelity $F(\chi,\omega)$
ranges between 0 and 1, $F(\chi,\omega)=1$ if and only if
$\chi=\omega$. The measurement over system I in mixed state $\rho$
produces result $R$ with probability
\begin{equation*}
P(R\,|\,\rho)={\rm Tr}_{\scriptscriptstyle{\rm I}}
[\, \Pi_R \: \rho \,]
\!\ ,
\end{equation*}
where $\Pi_R$ is the corresponding projector. Let the supremum in Eq.
(\ref{fiddef}) is reached by purifications $|u'\rangle$ and
$|v'\rangle$. Because
\begin{equation*}
P(R\,|\,\chi)=
\langle u'|\,\Pi_R \otimes {\boldsymbol{1}}\,|u'\rangle
\ \ {\rm and} \ \
P(R\,|\,\omega)= \langle v'|\,\Pi_R \otimes
{\boldsymbol{1}}\,|v'\rangle
\!\ ,
\end{equation*}
where ${\boldsymbol{1}}$ is the identity operator, Eq. (\ref{ineq1})
then gives
\begin{equation}
\bigl| P(R\,|\,\chi) - P(R\,|\,\omega)
\bigr| \leq \sqrt{1-\bigl|\langle u'|v'\rangle\bigr|^2}
=\sqrt{1-F(\chi,\omega)}
\!\ .
\label{ineqprobmix}
\end{equation}
The last relation extends Eq. (\ref{ineqprob}) to the case of mixed
states.

The measurement of transition probability is an other main form of
experiment with a quantum system. In this case we for a time allow
the studied system I to interact unitarily with the auxiliary system
I{\thinspace}I{\thinspace}. The system I{\thinspace}I will cause the
system I to perform some transitions. The probability that after the
expiry of time $t$ the system I will have some property described by
projector $\Pi$ is equal to
\begin{equation}
{\rm Tr}_{\scriptscriptstyle{\rm I}}
[\,\Pi\, \rho (t)\,]
={\rm Tr}\,
[\,(\Pi\otimes{\boldsymbol{1}})\,\sigma(t)\,]
\!\ ,
\label{prabpi}
\end{equation}
where density operator
$\:\rho (t)={\rm Tr}_{\scriptscriptstyle{\rm II}}\sigma(t)\:$
of system I is the partial trace of density operator $\sigma(t)$ of
composite system "I+I{\thinspace}I" over system
I{\thinspace}I\thinspace. If in initial moment the system I resides
in pure state $|s\rangle$ and the system I{\thinspace}I resides in
pure state $|m\rangle$, then after the expiry of time $t$ the state
of composite system "I+I{\thinspace}I" is described by density
operator \begin{equation*}
\sigma_m^{(s)}(t)=|V_m^{(s)}(t)\rangle\langle V_m^{(s)}(t)|
\!\ ,
\end{equation*}
where vector
$\:|V_m^{(s)}(t)\rangle={\rm U}(t)\;|s\rangle\otimes|m\rangle\:$,
${\rm U}(t)$ is the evolution operator of composite system
"I+I{\thinspace}I". We then get that probability is
\begin{equation*}
{\rm Tr}\,
[\,(\Pi\otimes{\boldsymbol{1}})\,\sigma_m^{(s)}(t)\,]
=\langle
V_m^{(s)}(t)|\,\Pi\otimes{\boldsymbol{1}}\,|V_m^{(s)}(t)\rangle \!\ .
\end{equation*}
The unitarity of transformation ${\rm U}(t)$ implies that
$\:\langle
V_m^{(\phi)}(t)|V_m^{(\psi)}(t)\rangle=\langle\phi|\psi\rangle\:$,
and by the use of Eq. (\ref{ineqprob}) we then have
\begin{equation}
\left|\,
{\rm Tr}\,
[\,(\Pi\otimes{\boldsymbol{1}})\,\sigma_m^{(\phi)}(t)\,]
-{\rm Tr}_{{\scriptscriptstyle{{\rm I}}+{\rm II}}}
[\,(\Pi\otimes{\boldsymbol{1}})\,\sigma_m^{(\psi)}(t)\,]
\,\right|\leq
\sin\delta_{\phi\psi} \!\ .
\label{propprob}
\end{equation}
for two initial states $|\phi\rangle$ and $|\psi\rangle$ of system I.
However, in many experiments auxiliary system I{\thinspace}I is the
macroscopic system, and, generally, its initial state is mixed.
Let the initial state of system I{\thinspace}I is described by density
operator
\begin{equation*}
\varrho (0) = \sum_m \mu_m
|m\rangle\langle m| \!\ , \quad \sum_m \mu_m = 1 \!\ .
\end{equation*}
Then after the expiry of time $t$ composite system "I+I{\thinspace}I"
will reside in the state described by density operator
\begin{equation*}
\sigma^{(s)} (t) =
{\rm U}(t)\left(
\sum_m \mu_m |s\rangle\langle s| \otimes |m\rangle\langle m|
\right){\rm U}^{\dagger}(t) =
\sum_m \mu_m \sigma_m^{(s)}(t)
\!\ .
\end{equation*}
Applying property
$\,{\rm Tr}\,({\rm A}+{\rm B})={\rm Tr}\,{\rm A}+{\rm Tr}\,{\rm
B}\,$, the triangle inequality and Eq. (\ref{propprob}), we then get
\begin{equation}
\left|\,
{\rm Tr}\,
[\,(\Pi\otimes{\boldsymbol{1}})\,\sigma^{(\phi)}(t)\,]
-{\rm Tr}\,
[\,(\Pi\otimes{\boldsymbol{1}})\,\sigma^{(\psi)}(t)\,]
\,\right|\leq
\sin\delta_{\phi\psi} \!\ .
\label{propro}
\end{equation}
Taking $\,\Pi=|\varphi\rangle\langle\varphi|\,$, for probability
$w(\varphi|s;t)$ of transition of system I from state $|s\rangle$
to state $|\varphi\rangle$ in a time $t$ we get such an inequality:
\begin{equation}
\bigl|
w(\varphi|\phi;t) - w(\varphi|\psi;t)
\bigr| \leq
\sin\delta_{\phi\psi} \!\ .
\label{tranprob}
\end{equation}
Eqs. (\ref{ineqprob}) and (\ref{tranprob}) show that if the angle
between two states is small, then experimental manifestations of
these states are close to each other. Thus, the angle between
two pure states is the reasonable measure of their closeness.

\protect\section{Basic definitions}

Let us assume that the copying machine has as input $N$ particles,
each from which is prepared in state $|s\rangle$, and the
copying machine must output a fixed number $L=M+N$ of similar
prepared particles (ideally, number $L$ of particles, prepared each
in state $|s\rangle$). Input state $|s\rangle$ is in a set
${\mathfrak{A}}=\{|\phi\rangle,|\psi\rangle\}$ of two pure states,
which we would like to copy. If initial state of the copying machine
is described by vector $|m\rangle$, then a CM action is
\begin{equation}
\forall \
|s\rangle \in {\mathfrak{A}} \ {\bf :} \quad
|s^{\otimes N}\rangle\otimes|m\rangle \longmapsto
|V_m^{(s)}\rangle \!\ .
\label{abs}
\end{equation}
The unitarity of a copying transformation implies that
\begin{equation}
(\langle\phi|\psi\rangle)^N=
\langle V_m^{(\phi)}|V_m^{(\psi)}\rangle \!\ , \quad
\left(\cos\delta_{\phi\psi}\right)^N=
\cos\delta(V_m^{(\phi)},V_m^{(\psi)})
\!\ .
\label{abs1}
\end{equation}
Following to paper \cite{hillery}, we
act on the output $|V_m^{(s)}\rangle$ by projector
$\:|s^{\otimes L}\rangle\langle s^{\otimes L}|\otimes{\mathbf{1}}\:$,
where $\mathbf{1}$ is the identity operator. Let vector
$\,|s^{\otimes L}\rangle\otimes|q_m^{(s)}\rangle\,$ be a result of
this action, then the output can be expressed as
\begin{equation}
|V_m^{(s)}\rangle=
|s^{\otimes L}\rangle\otimes|q_m^{(s)}\rangle +
|{\perp}_m^{(s)}\rangle
\!\ .
\label{buz}
\end{equation}
The idempotency of the above projector
implies
\begin{equation}
\bigl\{|s^{\otimes L}\rangle\langle s^{\otimes L}|\otimes
{\mathbf{1}}\bigr\}
|{\perp}_m^{(s)}\rangle = 0
\label{perp}
\!\ .
\end{equation}
In general,
$\,|q_m^{(s)}\rangle\,$ and $\,|{\perp}_m^{(s)}\rangle\,$ are not
unit, but there is
\begin{equation}
\|\>\! |q_m^{(s)}\rangle \|^2 +
\|\:\! |{\perp}_m^{(s)}\rangle \|^2 = 1
\label{buz1}
\end{equation}
according to the unitarity. Quantity
$\:X_m^{(s)}=\|\:\!|{\perp}_m^{(s)}\rangle\|\:$
was introbuced by Hillery and Bu\v{z}ek \cite{hillery} as the size of
error of copying of state $|s\rangle$. We shall now consider a
relationship between $X_m^{(s)}$ and the deviation of the resulting
probability distribution from the desired probability distribution.
Let us introduce magnitude
\begin{equation}
\delta_m^{(s)} = \inf \bigl\{ \
\delta(V_m^{(s)},s^{\otimes L}\otimes k) \ \big| \
\langle k|k\rangle=1 \ \bigr\} \!\ .
\label{deltas}
\end{equation}
Using relation (\ref{perp}),
we see that the inner product of unit vectors $\,\langle
V_m^{(s)}|\,$ and $\>|s^{\otimes L}\rangle\otimes|k\rangle\>$ is
equal to $\>\langle q_m^{(s)}|k\rangle\>$. Because
$\,\|\>\!|k\rangle\|=1\,$, the Schwarz inequality gives
\begin{equation*}
\bigl| \langle q_m^{(s)}|k\rangle \bigr| \leq \|\>\!
|q_m^{(s)}\rangle \| \!\ ,
\end{equation*}
where the equality takes place if and only if
$\>|q_m^{(s)}\rangle=c\>|k\rangle\>$ for some complex number $\,c\,$.
The maximal value of the modulus of the inner product of two unit
vectors corresponds to the minimal value of angle
between these vectors, so that if stated in Eq. (\ref{deltas})
infimum is reached by vector $|k\rangle$, then unit vector
$|k\rangle$ and vector $|q_m^{(s)}\rangle$ are collinear. For
$\,\|\>\!|q_m^{(s)}\rangle\|\not=0\,$ let us define vectors
\begin{equation}
|k_m^{(s)}\rangle
\ {\mathop=^{{\rm def}}} \
|q_m^{(s)}\rangle\bigm/\|\>\!|q_m^{(s)}\rangle\|
\ \ \ {\rm and} \ \ \
|{Id\,}_m^{(s)}\rangle
\ {\mathop=^{{\rm def}}} \
|s^{\otimes L}\rangle\otimes|k_m^{(s)}\rangle
\label{ideal}
\!\ .
\end{equation}
Stated in Eq. (\ref{deltas}) infimum is reached for each vector
$\,|k\rangle=u\,|k_m^{(s)}\rangle\,$ with complex unit $u\,$,
and $\delta_m^{(s)}$ is angle between unit vectors
$\,|V_m^{(s)}\rangle\,$ and $\,|{Id\,}_m^{(s)}\rangle\,$. Let
Hermitian operator ${\rm A}$ describes some observable for particle 1.
Its measurement over particle in state $|s\rangle$ produces result
$a$ with probability
$\:p(a|s)=\langle
s|\,\Pi_a|s\rangle\:$,
where $\Pi_a$ is the
corresponding projector. Consider now this observable for composite
system "1+$\cdots$+L+CM". In accordance with Eq. (\ref{prabpi}), the
measurement of such an observable over system "1+$\cdots$+L+CM" in
pure state $|V\rangle$ gives result $a$ with probability
\begin{equation*}
P(a\; {\rm for}\; 1\,|\,V)= \langle
V|\:\Pi_a\!\otimes{\mathbf{1}}^{\otimes L} |V\rangle \!\ ,
\end{equation*}
where $\:\Pi_a\otimes{\mathbf{1}}^{\otimes L}\:$ is the
projector on the corresponding subspace of the composite system state
space. In a similar manner, the expression for measurement of
observable for particle $j$ is obtained. For state
$|{Id\,}_m^{(s)}\rangle$ the probability of outcome $a$ is
\begin{equation}
P(a\; {\rm for}\;
j\,|\, {Id\,}_m^{(s)})= \langle s|\,\Pi_a|s\rangle =p(a|s) \!\ ,
\label{perfprob}
\end{equation}
where $\:j=1,\ldots,L\:$ and $s=\phi,\psi$. Thus,
$|{Id\,}_m^{(s)}\rangle$ corresponds to the ideal output. In line
with Eqs. (\ref{deltas}) and (\ref{ideal}) we have
\begin{equation}
\cos\delta_m^{(s)}=
\bigl|\langle V_m^{(s)}|{Id\,}_m^{(s)}\rangle
\bigr|=\bigl|\langle q_m^{(s)}|k_m^{(s)}\rangle\bigr|=
\|\>\! |q_m^{(s)}\rangle\|
\!\ .
\label{qms}
\end{equation}
Then Eq. (\ref{buz1}) gives
$\>\|\:\!|{\perp}_m^{(s)}\rangle\|=\sin\delta_m^{(s)}\>$.
As Eqs. (\ref{ineqprob}) and (\ref{perfprob}) and
definition of $X_m^{(s)}$ show,
\begin{equation}
\left| P(a\;
{\rm for}\; j\,|\, V_m^{(s)}) - p(a|s) \right| \leq X_m^{(s)} \!\ ,
\label{derpro}
\end{equation}
i.e. magnitude $X_m^{(s)}$ characterizes as a whole the
deviation of the resulting probability distribution from the desired
probability distribution. Sum $\>X_m^{(\phi)}+X_m^{(\psi)}\>$
evaluates the total error of copying of set $\mathfrak{A}$,
when the initial state of the copying machine is described by vector
$|m\rangle$.

{\bf Definition 1}
{\it The size
$\:AE({\mathfrak{A}})=X_m^{(\phi)}+X_m^{(\psi)}\:$
is the absolute error of copying for set
${\mathfrak{A}}=\{|\phi\rangle,|\psi\rangle\}$.}

However, this criterion loses sight of closeness of states
$|\phi\rangle$ and $|\psi\rangle$. To understand this better we
shall argue in the form of a game. Fixed parameters are the set
${\mathfrak{A}}=\{|\phi\rangle,|\psi\rangle\}$ of two non-orthogonal
states, the number $N$ of input qubits and the number $L$ of output
qubits. Two persons called Alice and Clare play the game.
(Following to paper \cite{werner}, we shall call the paradigmatical
cloner Clare.) Both players know the game parameters. At first,
Alice chooses one state from the set ${\mathfrak{A}}$ without
Clare's knowledge. The next Alice's move is to prepare each of $N$
qubits in the chosen state. Then Alice sends the prepared
qubits to Clare. Clare's step is to make $L>N$ qubits of the given
$N$ qubits. The next Clare's step is to guess the Alice's choice by
measurement made on the output of the copying machine. The game is
repeated. How many chances has Clare?

Let us take that states $|\phi\rangle$ and $|\psi\rangle$ are
sufficiently close to each other. Then the lower bound on the
absolute error is close to 0. Next, both the upper bounds on the
global fidelity and the local fidelity are close to 1. All these
criteria assert that copying process can be made near to the
ideality. Clare knows both the ideal outputs corresponding
to choise of $|\phi\rangle$ and choise of $|\psi\rangle$
respectively. She will compare given output to
this ideal output and to that one. At first sight it seems that
Clare can simply recognize the chosen state. It would be
a rashness to think so. Indeed, the closeness of states
$|\phi\rangle$ and $|\psi\rangle$ implies certain closeness of the
ideal outputs.  But if so, is Clare able to decide that given output
should be related to this ideal output and not to that one?  How are
Clare's chances dependent on the game parameters?

To express this in quantitative form we should use some measure of
closeness for states $|{I}_m^{(\phi)}\rangle$ and
$|{I}_m^{(\psi)}\rangle$. Since according to Eq. (\ref{ineqprob})
$$
\left| P(R\,|\, {Id\,}_m^{(\phi)}) -
P(R\,|\, {Id\,}_m^{(\psi)}) \right| \leq
\sin\delta({Id\,}_m^{(\phi)},{Id\,}_m^{(\psi)}) \!\ ,
$$
the quantity $\>\sin\delta({I}_m^{(\phi)},{I}_m^{(\psi)})\>$
provides such a measure. It stands to reason, this quantity is
depending on similarity of states $|\phi\rangle$ and
$|\psi\rangle$. Let us take that
$\>\sin\delta({I}_m^{(\phi)},{I}_m^{(\psi)})\>$ is small.  Is Clare
willing to decide that given output $|V_m^{(s)}\rangle$
should be related to ideal output $|{Id\,}_m^{(\phi)}\rangle$ and not
to $|{I}_m^{(\psi)}\rangle$ ? The closeness of $|V_m^{(\phi)}\rangle$
to $|{I}_m^{(\phi)}\rangle$ is measured by
$X_m^{(\phi)}=\sin\delta_m^{(\phi)}$, the closeness of
$|V_m^{(\psi)}\rangle$ to $|{I}_m^{(\psi)}\rangle$ is measured by
$X_m^{(\psi)}=\sin\delta_m^{(\psi)}$.
It is not without significance that
$\>\sin\delta({I}_m^{(\phi)},{I}_m^{(\psi)})\>$ is size of the
same kind. Therefore, it is advisable to compare
the absolute error with pointed quantity.

{\bf Definition 2}
{\it The relative error of $N\to L$ copying for set
${\mathfrak{A}}=\{|\phi\rangle,|\psi\rangle\}$ is}
\begin{equation}
RE({\mathfrak{A}}) \ {\mathop=^{{\rm def}}} \
AE({\mathfrak{A}})\,\Big/\sin\delta({Id\,}_m^{(\phi)},{Id\,}_m^{(\psi)})
\!\ .
\label{redef}
\end{equation}

We shall now derive the angle relations, from which bounds on the
errors are simply obtained. In order to be rid of bulky
expressions we shall below use the notation
\begin{equation}
\delta_N=\delta(\phi^{\otimes N},\psi^{\otimes N}) \!\ .
\end{equation}
Recall that the spherical triangle inequality holds \cite{rastegin}:
\begin{equation}
\delta(X,Y)\leq
\delta(X,Z)+\delta(Y,Z)
\!\ .
\label{cosplu}
\end{equation}
Using Eq. (\ref{cosplu}) twice, we have
\begin{equation}
\delta({Id\,}_m^{(\phi)},{Id\,}_m^{(\psi)}) \leq
\delta_m^{(\phi)} + \delta_m^{(\psi)} +
\delta(V_m^{(\phi)},V_m^{(\psi)}) \!\ .
\label{condit}
\end{equation}
In accordance with the Schwarz inequality, there is
$$
\bigl|\langle {Id\,}^{(\phi)}|
{Id\,}^{(\psi)}\rangle
\bigr|=\bigl|\langle \phi|\psi\rangle \bigr|^L \>
\bigl|\langle k^{(\phi)}|k^{(\psi)}\rangle \bigr|
\leq \bigl|\langle \phi|\psi\rangle \bigr|^L \!\ ,
$$
whence we obtain $\:\delta({Id\,}_m^{(\phi)},{Id\,}_m^{(\psi)})
\geq\delta_L\:$. Therefore, $\:\delta_L\leq
\delta_m^{(\phi)}+\delta_m^{(\psi)}+\delta(V_m^{(\phi)},V_m^{(\psi)})\:$,
or simply
\begin{equation}
\delta_m^{(\phi)} +
\delta_m^{(\psi)} \geq
\delta_L-\delta_N
\label{keyunit}
\end{equation}
in line with Eq. (\ref{abs1}). Since
$\,0\leq\bigl|\langle\phi|\psi\rangle\bigr|\leq1\,$ and $N<L$, there
is $\,\delta_N\leq\delta_L\,$. Eqs. (\ref{condit}) and
(\ref{keyunit}) contain the restrictions imposed by the laws of the
quantum theory. In particular, the ones allow to derive the lower
bounds on both the relative error and absolute error.

\protect\section{Lower bounds on the absolute and relative errors}

In this section the lower bounds on the absolute and relative errors
will be obtained. In order to minimize $RE({\mathfrak{A}})$ the
quantity $\>\sin\delta({Id\,}_m^{(\phi)},{Id\,}_m^{(\psi)})\>$ must
be as increased as possible. We shall individually consider two
cases:
\begin{itemize}
\item[(i)]{$\>\delta_m^{(\phi)}+\delta_m^{(\psi)}+
\delta_N\leq\pi/2\>$,}
\item[(ii)]{$\>\delta_m^{(\phi)}+\delta_m^{(\psi)}+
\delta_N>\pi/2\>$.}
\end{itemize}
Using Eqs. (\ref{abs1}) and (\ref{condit}), for the case (i) we have
\begin{equation}
\sin\delta({Id\,}_m^{(\phi)},{Id\,}_m^{(\psi)})\leq
\sin(\delta_m^{(\phi)}+\delta_m^{(\psi)}
+\delta_N) \!\ .
\end{equation}
In addition, there is (see trigonometric formula for sine of sum 
\cite{handbook})
\begin{equation}
\sin\delta_m^{(\phi)}+\sin\delta_m^{(\psi)}\geq
\sin(\delta_m^{(\phi)}+\delta_m^{(\psi)})
\label{epsilbn}
\end{equation}
By the two last inequalities,
\begin{equation}
RE({\mathfrak{A}})\geq
\cos\delta_N -
\sin\delta_N
\cot(\delta^{(\phi)}+\delta^{(\psi)}+\delta_N) \!\ .
\label{cot}
\end{equation}
It must be stressed that the equality in Eq. (\ref{epsilbn})
is necessary for the equality in Eq. (\ref{cot}). We want
minimizing the right-hand side of Eq. (\ref{cot}) in the interval
$\:\delta_L\leq\delta_m^{(\phi)}+\delta_m^{(\psi)}+\delta_N\leq\pi/2\:$
established by Eq. (\ref{keyunit}) and the case (i)
condition. The minimum is reached at the left boundary
point of the above interval. In fact, the right-hand side of Eq.
(\ref{cot}) increases as the cotangent decreases, and the cotangent
is a decreasing function of one's argument. Therefore, in the case
(i)
\begin{equation}
RE({\mathfrak{A}})\geq
\sin(\delta_L - \delta_N)\,/
\sin\delta_L \!\ .
\label{icasmin}
\end{equation}
Further, in the case (i) inequality
$\,AE({\mathfrak{A}})\geq\sin(\delta_L - \delta_N)\,$ holds.
In the case (ii)
$\>RE({\mathfrak{A}})\geq\sin\delta_m^{(\phi)}+\sin\delta_m^{(\psi)}\>$,
because $\>\sin\delta({Id\,}_m^{(\phi)},{Id\,}_m^{(\psi)})\leq1\>$.
Next, the case (ii) condition can be separated into two alternatives,
$\>\pi/2-\delta_N<
\delta_m^{(\phi)}+\delta_m^{(\psi)}\leq\pi/2\>$
and $\>\pi/2<\delta_m^{(\phi)}+\delta_m^{(\psi)}\leq\pi\>$. The first
alternative contains
$$
RE({\mathfrak{A}})\geq\sin\bigl(\delta_m^{(\phi)}+\delta_m^{(\psi)}\bigr)
\geq\cos\delta_N \!\ .
$$
In the second alternative the conditions
$\,\delta_m^{(s)}\leq\pi/2\,$ and
$\>\pi/2<\delta_m^{(\phi)}+\delta_m^{(\psi)}\leq\pi\>$ ensure
$\>\sin\delta_m^{(\phi)}+\sin\delta_m^{(\psi)}\geq1\>$.
So, in the case (ii)
$\>RE({\mathfrak{A}})\geq\cos\delta_N\>$ and
$\>AE({\mathfrak{A}})\geq\cos\delta_N\>$. To sum up, we see that
the lower bound on the relative error is given by the right-hand side
of Eq. (\ref{icasmin}). Designating $\,z=\cos\delta_{\phi\psi}\,$,
hence $\,\cos\delta_L=z^L\,$ and $\,\cos\delta_N=z^N\,$, Eq.
(\ref{icasmin}) can be rewritten as
\begin{equation}
RE({\mathfrak{A}})\geq
F(z|N,L)
\ {\mathop=^{{\rm def}}} \
z^N - z^L
\sqrt{(1-z^{2N})\big/(1-z^{2L})} \!\ .
\label{theore1}
\end{equation}

In addition, we have established that
$\,A({\mathfrak{A}})\geq\sin(\delta_L - \delta_N)\,$. This inequality
can be reformulated as the following inequality which improves the
results of Ref. \cite{hillery}:
\begin{equation}
AE({\mathfrak{A}}) \geq z^N \sqrt{1-z^{2L}} -
z^L \sqrt{1-z^{2N}} \!\ .
\label{theore2}
\end{equation}

Note that the derived lower bounds remain valid in the case, when the
initial state of the copying machine is mixed. This extension has
meaning, because the copying machine, most likely, be the macroscopic
system. If the initial state of the copying machine is described
by density operator
\begin{equation}
\varrho = \sum_m \mu_m
|m\rangle\langle m| \!\ , \quad \sum_m \mu_m = 1 \!\ ,
\label{denop}
\end{equation}
then after the copying procedure the composite system
"1+$\cdots$+L+CM" will reside in the state described by density
operator
\begin{equation}
\sigma^{(s)} = \sum_m \mu_m
|V_m^{(s)}\rangle\langle V_m^{(s)}| \!\ .
\label{afden}
\end{equation}
Then the size of the error of copying of state $|s\rangle$ is defined
in a reasonable way as
\begin{equation}
X^{(s)}= \sum_m \mu_m
X_m^{(s)} \!\ .
\label{sizmix}
\end{equation}
Then the measurement of particle observable over particle 1 produces
result $a$ with probability
\begin{equation*}
P(a\; {\rm for}\; 1\,|\,\sigma^{(s)})=
{\rm Tr} \left[ \left(
\Pi_a\!\otimes{\mathbf{1}}^{\otimes L}\right)
\sigma^{(s)} \right]
\!\ ,
\end{equation*}
The expression for particle $j$ is simply obtained by obvious
changes. Using Eqs. (\ref{derpro}) and (\ref{sizmix}), we then get by
a way, which is similar to reason for (\ref{propro}), such an
inequality
\begin{equation}
\left| P(a\; {\rm for}\; j\,|\,
\sigma^{(s)}) - p(a|s) \right| \leq X^{(s)} \!\ ,
\label{promix}
\end{equation}
where $\:j=1,\ldots,L\:$ and $s=\phi,\psi$. In this case the absolute
error $\,AE({\mathfrak{A}})=X^{(\phi)}+X^{(\psi)}\,$, the relative
error
\begin{equation}
RE({\mathfrak{A}}) = \sum_m
\mu_m  \, \frac{X_m^{(\phi)}+X_m^{(\psi)}}
{\sin\delta({Id\,}_m^{(\phi)},{Id\,}_m^{(\psi)})}
\!\ .
\label{redefmix}
\end{equation}

Note that the lower bounds given by theorems 1 and 2 are tightest.
Indeed, we shall below describe the cloner that reaches the ones.
For example, $F(z|N,L)$ is plotted as function of $z$ for $N=1$ and
five values of $L$ in Fig.~1. In the greater part of interval
$z\in[0;1]$ the function increases and only in the vicinity of the
right boundary point the one becomes decreasing. The maximum of
function and the limiting value $\bigl(1-\sqrt{N/L}\,\bigr)$ as
$z\to1$ are values of the same order. So, quantity
$\bigl(1-\sqrt{N/L}\,\bigr)$ can estimate our possibilities for
the state-dependent $N\to L$ copying. If the relative number of
actual copies is small, i.e. $N/L\approx1$, then for all $z\in[0;1]$
the lower bound on the relative error is also small. Theoretically,
in this case we can attain the good quality of state-dependent
cloning. Conversely, if the relative number of actual copies is close
to 1, i.e. $N/L\ll1$, then the relative error will be perceptible
(except almost orthogonal states). In the limit $\,L\to\infty\,$ we
have $\,F(z|N,L)\to z^N\,$. Thus, there is, in general, a tradeoff
between a quality of copies and the relative number of actual copies.
As to the game played by Alice and Clare, we note the following. If
for given parameters of the game the quantity $F(z|N,L)$ is value of
order 1 then Clare hardly has chances. Too strong a closeness of
ideal outputs will prevent her from guessing Alice's choise.

\protect\section{Optimal asymmetric cloner}

We shall now describe the asymmetric state-dependent cloner reaching
the presented lower bounds. In principle, both lower bounds given by
Eqs. (\ref{theore1}) and (\ref{theore2}) can be reached without
ancilla. Then a unitary operator ${\rm U}$ acts on the Hilbert space
of $L$ qubits:
$$
|V^{(s)}\rangle={\rm U}\left\{
|s^{\otimes N}\rangle\otimes
|0^{\otimes M}\rangle\right\}
$$
for $s=\phi,\psi$. The ideal output
$\,|{Id\,}^{(s)}\rangle=|s^{\otimes L}\rangle\,$, and
$\>\sin\delta({Id\,}^{(\phi)},{Id\,}^{(\psi)})=\sqrt{1-z^{2L}}\>$.
The equality in Eq. (\ref{keyunit}) is necessary to
minimize the relative error. Recall that the equality in Eq.
(\ref{cosplu}) holds only if the triplet is coplanar \cite{rastegin}.
Therefore the equality in Eq. (\ref{keyunit}) holds
only if both final states $|V^{(\phi)}\rangle$ and
$|V^{(\psi)}\rangle$ lie in plane $\,{\rm span}\{|\phi^{\otimes
L}\rangle,|\psi^{\otimes L}\rangle\}\,$. This is also necessary to
maximize the global fidelity \cite{brass,macchi}. Because unitary
operations preserve angles, we have
\begin{equation}
\delta(V^{(\phi)},V^{(\psi)})=
\delta(\psi^{\otimes N}\otimes0^{\otimes M},
\psi^{\otimes N}\otimes0^{\otimes M}) \!\ . \label{vdnm}
\end{equation}
If states $|\phi\rangle$ and $|\phi\rangle$ are not orthogonal or
identical then angle
$\,\delta(\phi^{\otimes L},\psi^{\otimes L})\,$
is larger than the right-hand side of Eq. (\ref{vdnm}), and
the ideal copying is impossible. In fact, it is impracticable that
angle between
$\:|\phi^{\otimes N}\otimes0^{\otimes M}\rangle\:$ and
$\:|\psi^{\otimes N}\otimes0^{\otimes M}\rangle\:$ should be properly
increased. To superpose the plane
$\:{\rm span}\{|\phi^{\otimes N}\otimes
0^{\otimes M}\rangle,|\psi^{\otimes N}\otimes0^{\otimes
M}\rangle\}\:$ onto the plane $\,{\rm span}\{|\phi^{\otimes
L}\rangle,|\psi^{\otimes L}\rangle\}\,$
by rigid rotation ${\rm U}$ is at most that we can
achieve. The transformation with characteristics
\begin{align}
& {\rm span}\{|V^{(\phi)}\rangle,|V^{(\psi)}\rangle\}
={\rm span}\{|\phi^{\otimes L}\rangle,|\psi^{\otimes L}\rangle\}
\!\ , \label{symclo1} \\
& \delta^{(\phi)}=\delta^{(\psi)} = (\delta_L
-\delta_N)\big/2
\label{symclo2}
\end{align}
is the optimal 'global' cloner constructed
in Refs. \cite{brass} (for the $1\to2$ case) and \cite{macchi}. This
cloner produces equal errors for both states $|\phi\rangle$ and
$|\psi\rangle$. The absolute error
$\,AE_{S}({\mathfrak{A}})=2\sin\,[(\delta_L-\delta_N)/2]\,$. Using
the standard trigonometric formulae we find that the relative error
$$
RE_{S}({\mathfrak{A}})=\sqrt{2}\,\left[
\:\frac{1-z^{N+L}}{1-z^{2L}} - \sqrt{\frac{1-z^{2N}}{1-z^{2L}}}
\:\right]^{1/2} \!\ .
$$
This cloner does not reach the equality in
Eq. (\ref{epsilbn}) (except when states $|\phi\rangle$ and
$|\psi\rangle$ are orthogonal or identical). Therefore, the optimal
'global' cloner minimizes neither the relative error nor the absolute
error.

We shall now propose an asymmetric cloner optimizing the relative
error. Such a optimal asymmetric state-dependent cloner is defined by
\begin{align}
& {\rm span}\{|V^{(\phi)}\rangle,|V^{(\psi)}\rangle\}
={\rm span}\{|\phi^{\otimes L}\rangle,|\psi^{\otimes L}\rangle\}
\!\ , \label{asclo1} \\
& \delta^{(\phi)}=0 \wedge
\delta^{(\psi)} = \delta_L
-\delta_N \!\ .
\label{asclo2}
\end{align}
This cloner makes the ideal copying of one from pair
$\mathfrak{A}$ of prescribed states. Both the equality in Eq.
(\ref{keyunit}) and the equality in Eq. (\ref{epsilbn}) are reached.
Therefore, for the cloner defined by Eqs. (\ref{asclo1}) and
(\ref{asclo2}) the relative error $\,RE_A({\mathfrak{A}})=F(z|N,L)\,$
and the absolute error $AE_A({\mathfrak{A}})$ is equal to the
right-hand side of Eq. (\ref{theore2}). In other words, the optimal
asymmetric state-dependent cloner minimizes both the relative and
absolute errors.

Thus, if states $|\phi\rangle$ and $|\psi\rangle$ are not orthogonal
or identical then
$\:RE_A({\mathfrak{A}})<RE_S({\mathfrak{A}})\:$. Is this distinction
significant? In order to study the question we consider the relative
value of difference between $RE_S({\mathfrak{A}})$ and
$RE_A({\mathfrak{A}})$, that is
\begin{equation}
{\rm f}\>\!(z|N,L)
=\bigl\{RE_S({\mathfrak{A}})-RE_A({\mathfrak{A}})\bigr\}
\big/RE_A({\mathfrak{A}})
\!\ .
\label{fznl}
\end{equation}
For example, ${\rm f}\>\!(z|N,L)$ is plotted as
function of $z$ for $N=1$ and five values of $L$ in Fig.~2. One
sees that the distinction between $RE_S({\mathfrak{A}})$ and
$RE_A({\mathfrak{A}})$ reaches several interest and becomes
perceptible as the part of actual copies increases. Quantity
${\rm f}\>\!(z|N,L)$ also illustrates the distinction between
$AE_S({\mathfrak{A}})$ and $AE_A({\mathfrak{A}})$. Thus, the cloner
defined by Eqs. (\ref{asclo1}) and (\ref{asclo2}) is not
insignificant.

\protect\section{Conclusion}

We have studied new optimality criterion for the state-dependent
$N\to L$ cloning. We have beforehand presented a few useful
inequalities. Using physical reasons, the notion of the relative
error has been then introduced. We have found that optimizing the
relative error principally differs from optimizing other
criteria. The tightest lower bounds on both the relative error and the
absolute error have been obtained. The ones depend on the number $N$
of input qubits, the number $L$ of input qubits and argument $z$
that is the modulus of the inner product of states to be
copied. We have described the optimal asymmetric state-dependent
cloner that minimizes both the relative error and the absolute error.
For studying with restect to 'fidelity' viewpoint, optimizing
the relative error is complementary rather than competitive.
Thus, the study of the relative error has allowed to complement a
portrait of the state-dependent $N\to L$ cloning.

\newpage

\begin{figure}[t!] %%%%%%%%%%%%%%%%%%%%%%%%%%%%%%%%
\vskip -20mm
\centering{\mbox{\epsfig{file=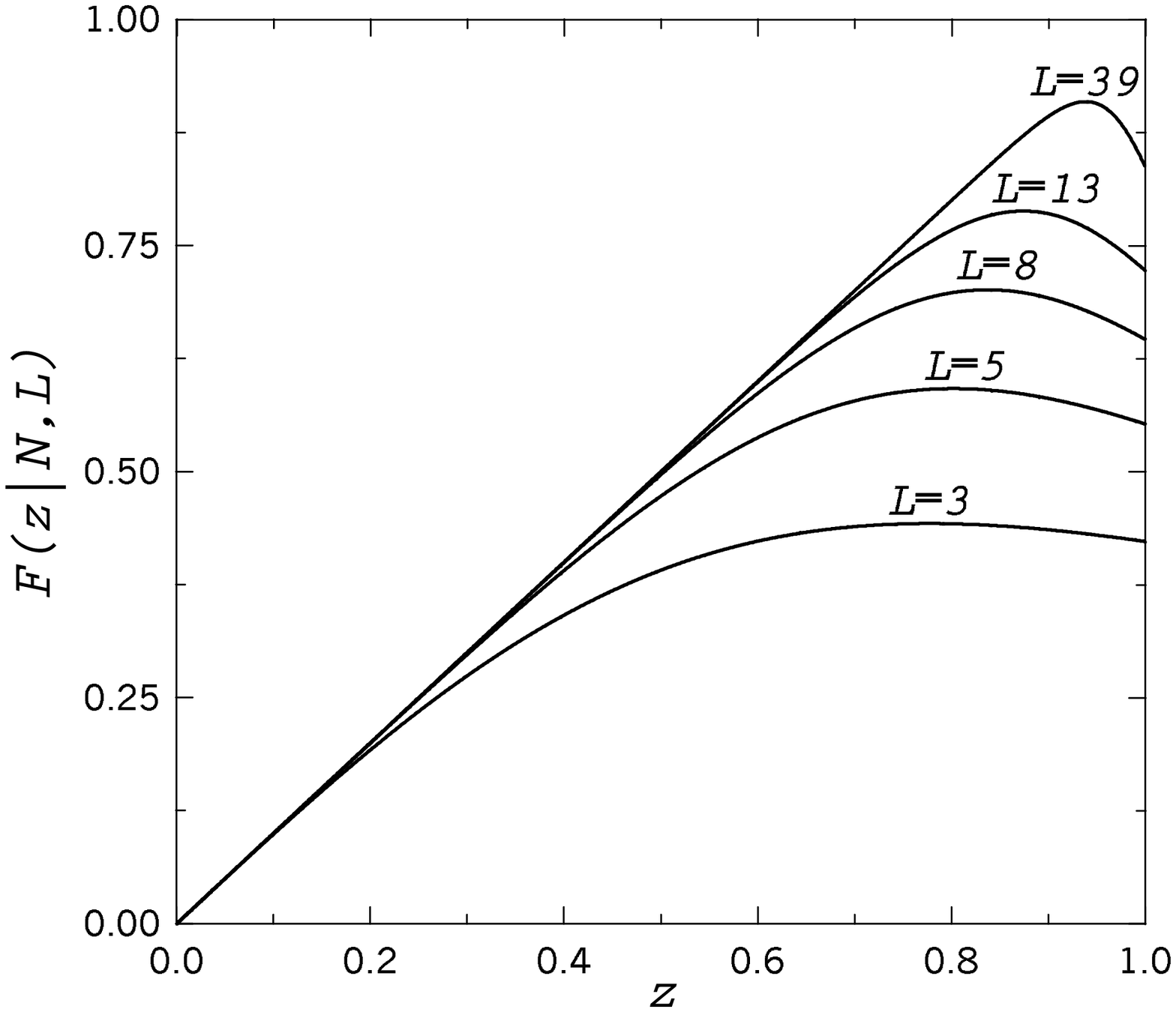,width=14.cm}}}
\vskip 5mm
\caption{The function defined by Eq. (\ref{theore1}) for $N=1$
	    and five values of $L$, namely $L=3,5,8,13,39$.}
\end{figure} %%%%%%%%%%%%%%%%%%%%%%%%%%%%%%%%%%%%%%

\newpage

\begin{figure}[t!] %%%%%%%%%%%%%%%%%%%%%%%%%%%%%%%%
\vskip -20mm
\centering{\mbox{\epsfig{file=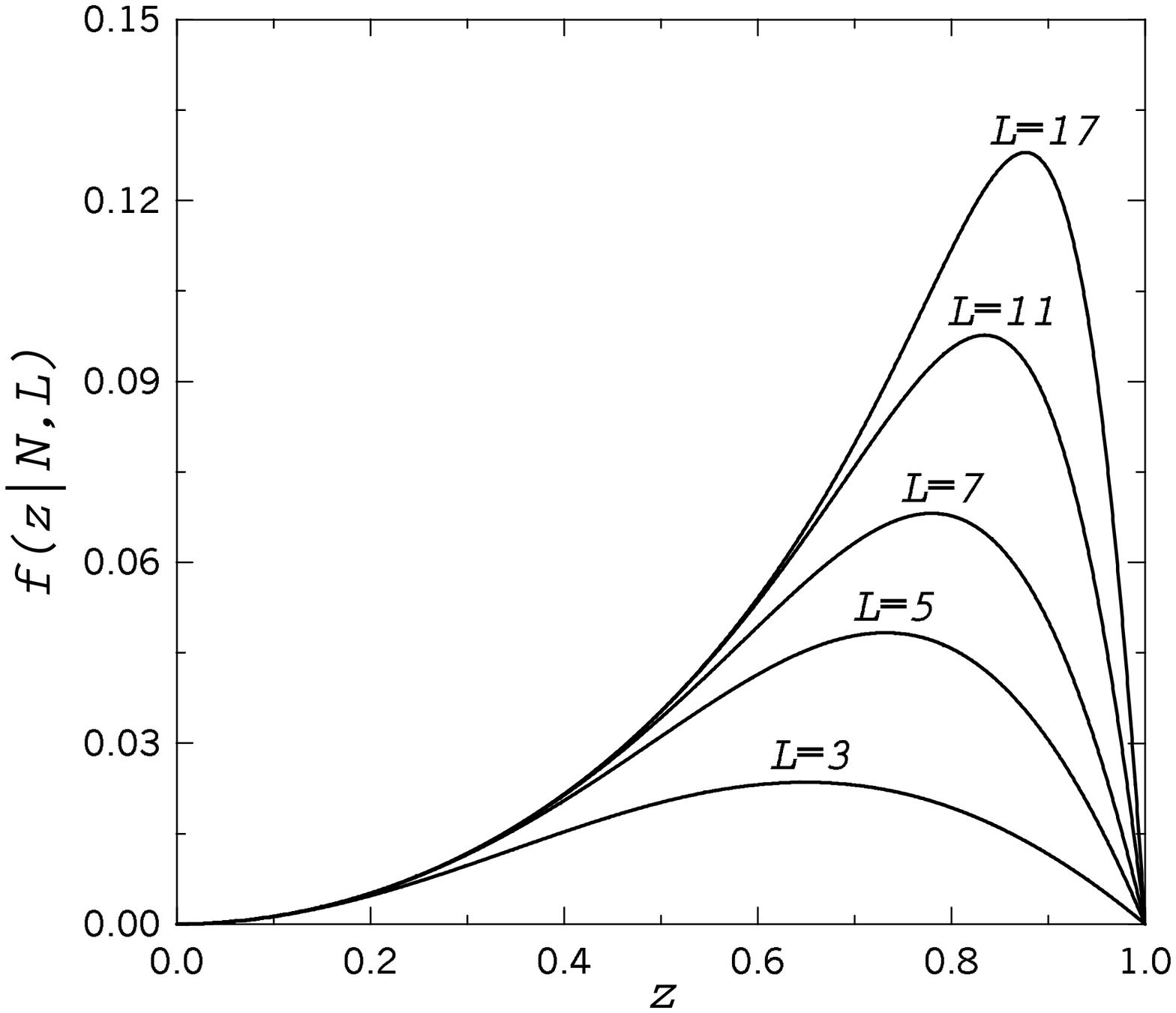,width=14.cm}}}
\vskip 5mm
\caption{The function defined by Eq. (\ref{fznl}) for $N=1$
	    and five values of $L$, namely $L=3,5,7,11,17$.}
\end{figure} %%%%%%%%%%%%%%%%%%%%%%%%%%%%%%%%%%%%%%


\begin{references}

\bibitem{wootters}%-----------------------------------------------
W.~K.~Wootters and W.~H.~Zurek, A single quantum cannot be cloned,
Nature {\bf 299}, 802--803 (1982)

\bibitem{barnum}%-----------------------------------------------
H.~Barnum, C.~M.~Caves, C.~A.~Fuchs, R.~Jozsa and B.~Schumacher,
Noncommuting mixed states cannot be broadcast, Phys. Rev. Lett. {\bf
76}, 2818--2821 (1996)

\bibitem{hillery}%-------------------------------------------------
M.~Hillery and V.~Bu\v{z}ek, Quantum copying: fundamental
inequalities, Phys. Rev. A {\bf 56}, 1212--1216 (1997)

\bibitem{brass}%--------------------------------------------------
D.~Bru{\ss}, D.P.~DiVincenzo, A.~Ekert, C.A.~Fusch, C.~Macchiavello,
J.A.~Smolin,  Optimal universal and state-dependent quantum cloning,
Phys. Rev. A {\bf 57}, 2368--2378 (1998), see also quant--ph/9705038

\bibitem{chefles}%------------------------------------------------
A.~Chefles and S.~M.~Barnett, Strategies and Networks for
State-Dependent Quantum Cloning, LANL report quant-ph/9812035

\bibitem{macchi}%------------------------------------------------
C.~Macchiavello, Bounds on the efficiency of cloning for two-state
quantum systems, J. Optics B {\bf 2}, 144--150 (2000)

\bibitem{rastegin}%-----------------------------------------------
A.~E.~Rastegin, Some bounds for quantum copying, LANL report
quant-ph/0108014

\bibitem{werner}%-----------------------------------------------
R.~F.~Werner, Optimal cloning of pure state, Phys. Rev. A {\bf 58},
1827--1832 (1998)

\bibitem{zanardi}%-----------------------------------------------
P.~Zanardi, A Note on Quantum Cloning in $d$ dimensions, LANL report
quant-ph/9804011

\bibitem{jozsa}%-----------------------------------------------
R.~Jozsa, Fidelity for mixed quantum states, J. Mod. Optics {\bf 41},
2315--2323 (1994)

\bibitem{handbook}%-----------------------------------------------
{\it Handbook of Mathematical Functions}, edited by M.~Abramovitz and
I.~A.~Stegun (National Bureau of Standards, Washington, 1964)

\end{references}
\end{document}